\documentclass[12pt]{article}

\usepackage{a4wide}
\usepackage[pdftex,usenames,dvipsnames]{color}
\usepackage{graphics}
\usepackage{amsfonts}
\usepackage{amssymb}
\usepackage{accents}

\newcommand{\nit}{\noindent}

\newcommand{\dsp}{\displaystyle}
\newcommand{\vs}[1]{\vspace{#1 ex}}
\newcommand{\hs}[1]{\hspace{#1 em}}
\newcommand{\bfr}{\begin{flushright}}
\newcommand{\efr}{\end{flushright}}
\newcommand{\bc}{\begin{center}}
\newcommand{\ec}{\end{center}}
\newcommand{\ben}{\begin{enumerate}}
\newcommand{\een}{\end{enumerate}}

\newcommand{\be}{\begin{equation}}
\newcommand{\ee}{\end{equation}}
\newcommand{\ba}{\begin{array}}
\newcommand{\ea}{\end{array}}
\newcommand{\ct}{\cite}
\newcommand{\bit}{\bibitem}

\newcommand{\bg}{\beta}
\newcommand{\gam}{\gamma}
\newcommand{\del}{\delta}

\newcommand{\thg}{\theta}

\newcommand{\lb}{\lambda}
\newcommand{\sg}{\sigma}

\newcommand{\Del}{\Delta}

\newcommand{\Og}{\Omega}

\newcommand{\cA}{{\cal A}}

\newcommand{\cM}{{\cal M}}

\newcommand{\sfm}{{\sf m}}

\newcommand{\lh}{\left(}
\newcommand{\rh}{\right)}
\newcommand{\ld}{\left.}
\newcommand{\rd}{\right.}

\newcommand{\slashed}{\hspace{-1.1ex}/}

\newcommand{\der}{\partial}

\newcommand{\sder}{\der\slashed}

\begin{document}

\pagestyle{plain}
\hfill{NIKHEF-2025-13}

\bc
{\bf \large Signature of Majorana neutrinos \\ ~ \\ in equal-sign {\boldmath$W$}-production by lepton scattering}
\vs{4}

{\large J.W.\ van Holten}
\vs{3}

Nikhef, Amsterdam NL \\
and \\
Leiden University. Leiden NL
\ec
\vs{3}

\nit
{\small {\bf Abstract} \\
The question whether standard-model neutrinos are Dirac or Majorana particles is presently undecided.
In principle it could be decided in favor of a Majorana signature if same-sign charged lepton scattering
would be observed to produce $W$-boson pairs, which is a lepton-number violating process. The cross 
section of this scattering process is computed; depending on the initial state the result gives access to
different components of the mass matrix. It could also shed light on physics beyond the standard 
model. }
\vs{3}

\nit
{\bf 1.\ Neutrinos and their signature}
\vs{1}

\nit
Neutrino oscillations provide proof of the existence of massive light neutrinos as part of the 
standard model; for an extensive review, see \ct{pdg:2020}. Whether these 
neutrino masses have a Dirac or Majorana signature is presently undecided. Its resolution offers 
a potential window on physics beyond the standard model. Active lines of investigation to measure
neutrino masses directly and to settle the question of the signature are $\bg$-decay experiments 
and the search for neutrinoless double-$\bg$ decay \ct{cuore:2022,katrin:2024}; 
for reviews, see \ct{dolinski:2019, formaggio:2021}. Neutrinoless double-$\bg$ decay is the process 
illustrated in fig.\,1, representing a nucleus in which two nucleons simultaneously decay by emitting 
a charged $W$-boson and exchange a Majorana neutrino $N$ to produce a pair of electrons 
\ct{schechter:1980}.
\bc
\vs{-2}
\scalebox{0.18}{\includegraphics{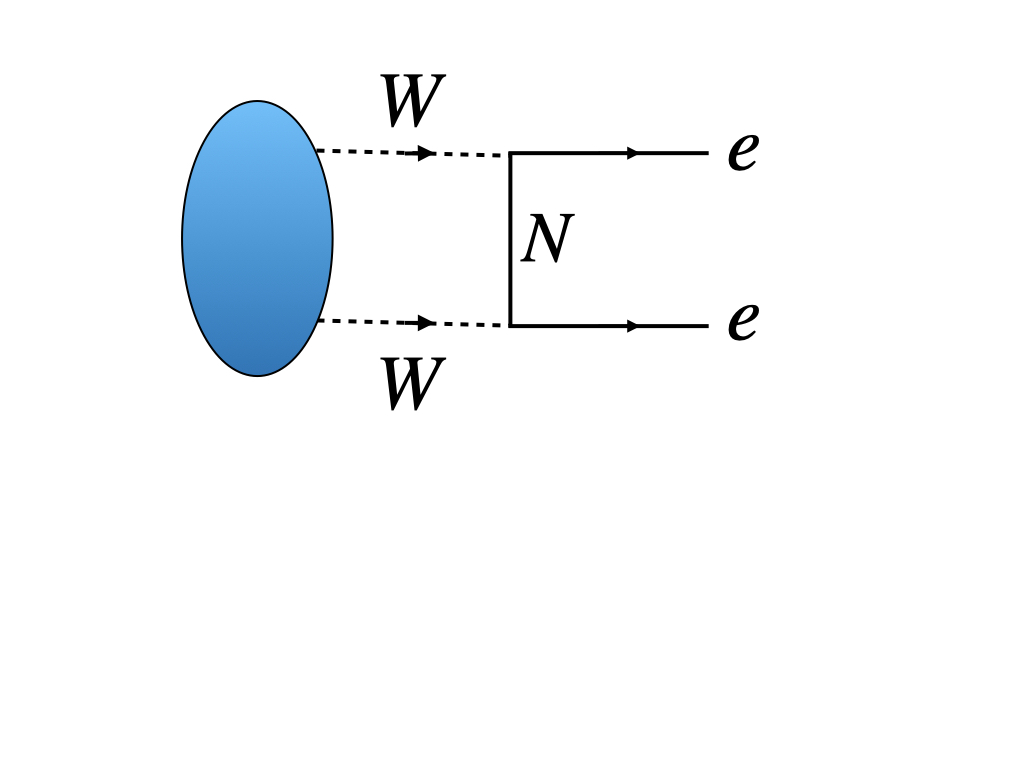}}
\vs{-11}

\footnotesize{Fig.\,1: Neutrinoless double-$\bg$ decay}
\ec
These experiments are extremely difficult due to complicating nuclear physics effects, and because
the amplitude for the process is proportional to the neutrino mass which is very small. 

\nit
Although the latter problem cannot be overcome, a much cleaner experiment is obtained in principle
from the reverse process: the scattering of two equal-sign charged leptons to create a pair of 
$W$-bosons by exchanging a Majorana neutrino \ct{london:1987,gluza:1995, greub:1996, kolodziej:1997}. 
Like neutrinoless double-$\bg$ decay this is a process which violates lepton number by $\Del L = 2$, a 
clear sign of the involvement of a Majorana particle. By scattering not only electrons but muons as well, 
such an experiment --if realizable-- could also provide access to other elements of the neutrino mass matrix 
for Majorana neutrinos. 
\vs{2}

\nit
{\bf 2.\ Action, amplitude and cross section}
\vs{1}

\nit
The part of the action of the of the standard model describing such processes in the mass base of
Majorana neutrino fields is 
\be
\ba{lll}
S_{EW} & = & \dsp{ \int d^4 x \left[ i \bar{E} \lh \sder + m_E \rh E - \frac{i}{2}\, N^T C \lh \sder N + m_N \rh N 
 - \frac{1}{2}\,F^{\mu\nu}(W^+)F_{\mu\nu}(W^-) \rd }\\
 & &\\
 & & \dsp{ \ld -\, M^2\, W^+ \cdot W^- + \frac{g_2}{2\sqrt{2}}\, \lh \bar{E}\, W\hs{-1}/^{\;\;\,-} (1 + \gam_5)\, U N
 - N^T C\, U^{\dagger} W\hs{-1}/^{\;\;\,+} (1 + \gam_5) E \rh  \right], }
\ea
\label{1}
\ee
where $T$ denotes transposition of spinor components, $C$ is the charge-conjugation matrix and 
$U$ is the unitary PNMS matrix mediating between mass and weak-interaction eigenstates of the neutrinos:
\[
\nu_{Li} = \frac{1}{2}\, (1 + \gam_5) \sum_j U_{ij} N_j.
\]
In the base $(E, N)$ of lepton fields the mass matrices of the leptons are taken to be diagonal: 
$m_E =$ diag$\,(m_e, m_{\mu}, m_{\tau})$ and $m_N =$ diag$\,(\mu_1,\mu_2,\mu_3)$. 

The relevant Feynman rules derived from this action are summarized in fig.\,2 below.\footnote{I
follow the conventions of  De Wit and Smith \ct{dewit:1986}.}
\bc
\vs{-1.7}
\scalebox{0.23}{\includegraphics{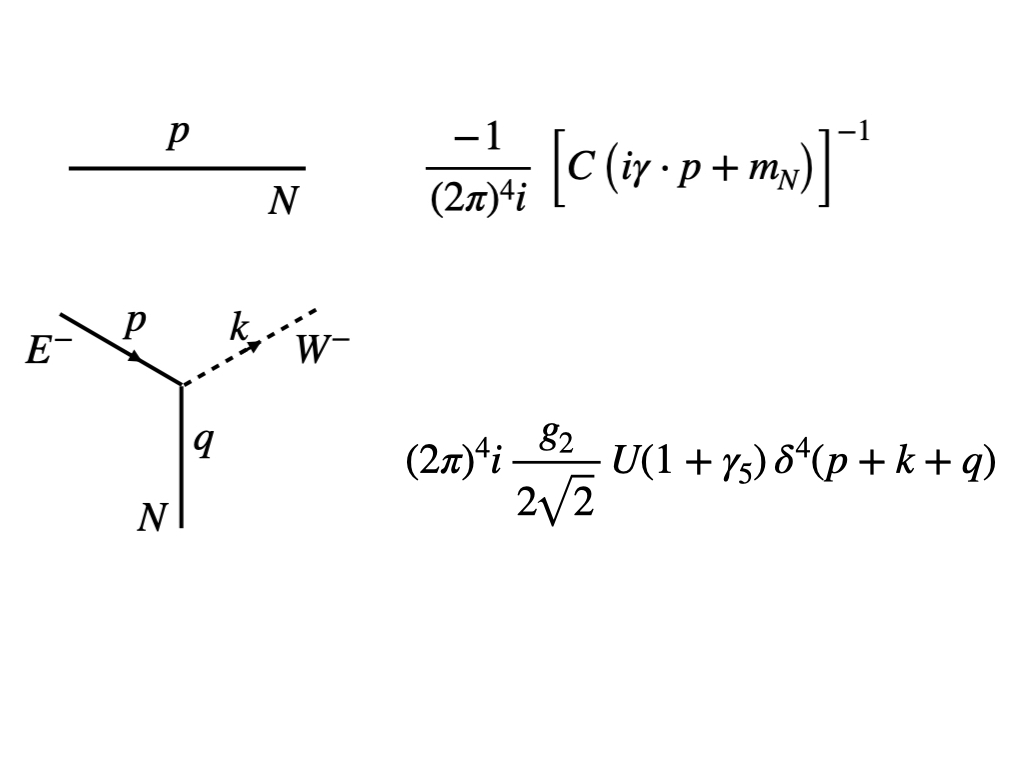}}
\vs{-6}

\footnotesize{Fig.\,2: Feynman rules for Majorana neutrinos}
\ec
The process of $W$-pair production by equal-sign lepton scattering: $E_i E_j \rightarrow WW$, 
is represented at tree level by the diagrams of fig.\,3.
\bc
\vs{-1.5}
\scalebox{0.2}{\includegraphics{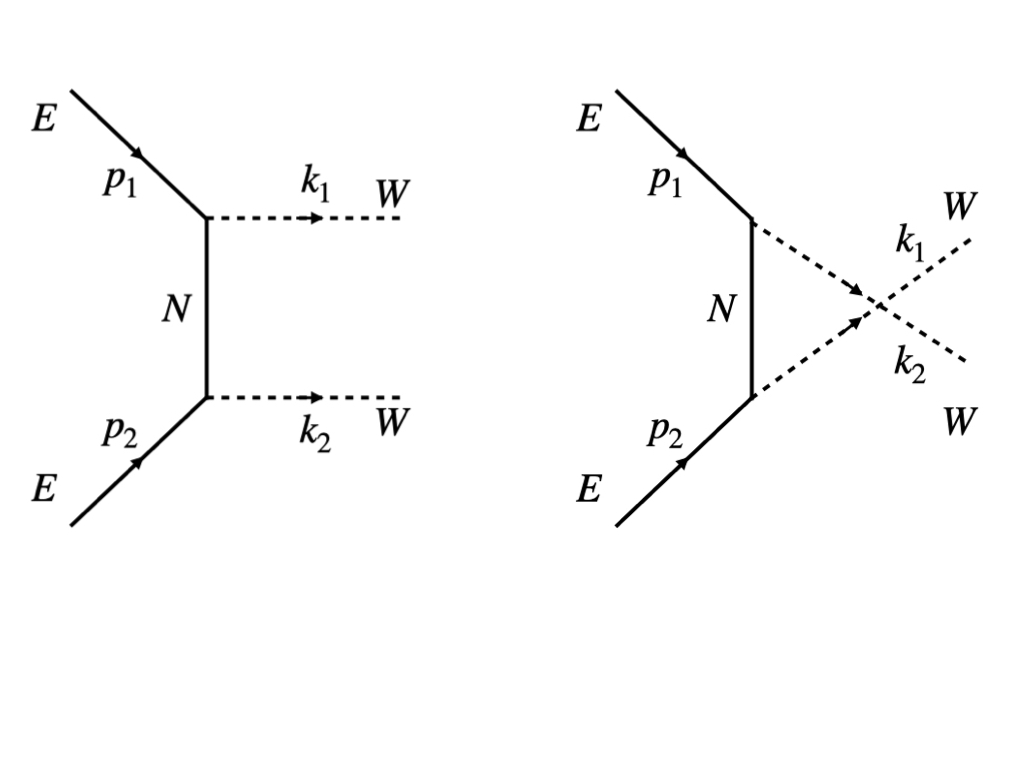}}
\vs{-7}

\footnotesize{Fig.\,3: $W$-pair production by charged lepton scattering}
\ec
The corresponding expression for the amplitude is
\be
\cA_{ij} = (2\pi)^4 i\, \cM_{ij}\, \del^4(p_1 + p_2 - k_1 - k_2),
\label{2}
\ee
where
\be
\cM_{ij} = - \frac{g_2^2}{2} \lh e(k_1) \cdot e(k_2) \rh 
 \sum_k \frac{\mu_k U_{ik} U_{jk}}{(p_1 - k_1)^2 + \mu_k^2}\, \bar{u}_i(p_1) (1 - \gam_5) u_j^c(p_2).
\label{3}
\ee
The notation $u^c = C\bar{u}^T$ is used for the charge-conjugate of an incoming spinor $u$.
As compared to the other masses and energies involved the neutrino masses are very small: 
$\mu^2_k \ll (m^2_i, m^2_j, M^2)$, their contribution to the denominator can be neglected. 
Introducing the complex mass matrix 
\[
\sfm_{Nij} = \sum_k \mu_k U_{ik} U_{jk},
\]
and using the standard notations $s = -(p_1+ p_2)^2$, $t = -(p_1 - k_1)^2$ the expression simplifies to
\be
\cM_{ij} = \frac{g_2^2\, \sfm_{Nij}}{2t}\, \lh e(k_1) \cdot e(k_2) \rh \bar{u}_i(p_1) (1 - \gam_5) u_j^c(p_2).
\label{4}
\ee
Using this result the differential cross section becomes
\be
\frac{d\sg}{d\Og_{CM}} = \thg(s - 4M^2) \sqrt{\frac{s(s - 4M^2)}{\lb(s,m_i^2,m_j^2)}}\, 
 \frac{g_2^4 |\sfm_{Nij}|^2}{256 \pi^2 s t^2}\,  \lh e(k_1) \cdot e(k_2) \rh^2 
 |\bar{u}_i(p_1) (1 - \gam_5) u_j^c(p_2)|^2,
\label{5}
\ee
with
\[
\lb(s,m_i^2,m_j^2) = s^2 - 2 s (m_i^2 + m_j^2) + (m_i^2 - m_j^2)^2. 
\]
As $s > 4M^2$ and the charged-lepton masses are small compared to the $W$-mass this 
reduces effectively to $\lb(s, m_i^2, m_j^2) \simeq s^2$. After averaging over the incoming lepton
polarizations using
\[
\sum_{pol} u(p) \bar{u}(p) = i p\slashed + m, \hs{2} \sum_{pol} u^c(p) \bar{u}^c(p) = i p\slashed - m,
\]
and summing over all polarizations of the $W$-bosons using
\[
\sum_{pol} e^{\mu}(k) e^{\nu}(k)  = \eta^{\mu\nu} + \frac{k^{\mu}k^{\nu}}{M^2},
\]
this gives
\be
\lh \frac{d\sg}{d\Og_{CM}} \rh_{unpol} =  \thg(s - 4M^2) \sqrt{1 - \frac{4M^2}{s}}\, 
 \lh 3 + \frac{s(s - 4M^2)}{4M^4} \rh \frac{g_2^4 |\sfm_{Nij}|^2}{256 \pi^2 t^2}.
\label{6}
\ee
Finally, integrating over all angles with $t$ in this approximation varying between
\[
t_{\pm} = M^2 - \frac{s}{2} \pm \frac{1}{2} \sqrt{s(s - 4M^2)},
\]
the total unpolarized cross section becomes:
\be
\ba{lll}
\sg^{unpol} & = & \dsp{ \thg(s - 4M^2)\, \frac{g_2^4 |\sfm_{Nij}|^2}{64 \pi M^4} \sqrt{1 - \frac{4M^2}{s}}
 \lh 3 + \frac{s(s - 4M^2)}{4M^4} \rh }\\
 & & \\ 
& = & \dsp{ \thg(s - 4M^2)\, \frac{G_F^2 |\sfm_{Nij}|^2}{2 \pi} \sqrt{1 - \frac{4M^2}{s}}
 \lh 3 + \frac{s(s - 4M^2)}{4M^4} \rh.}
\ea
\label{7}
\ee
Note that, although the cross section is proportional to the square of the neutrino mass matrix 
element, which is very small, at energies $\sqrt{s} \gg 2M$ it grows as $s^2/4M^4$. 
\vs{1}

\nit
{\bf 3.\ Discussion}
\vs{1}

\nit
In the standard model neutrino masses can arise by including a number of sterile neutrino species, 
or from effective low-energy interactions. Depending on the realization this allows for the light 
neutrinos of the standard model to be either of Dirac or Majorana type. Resolving this issue and 
determining neutrino mass matrix would provide a window on physics beyond the standard model. 
This is in particular the case if the scenario of the see-saw mechanism is realized, 
which could involve additional sterile neutrinos with Majorana masses in the range of 
grand unified models, but can at the same time explain the very small Majorana masses of 
the standard-model neutrinos \ct{gell-mann:1979, yanagida:1980}. 

Obviously these small standard-model masses make the cross section for $E_iE_j \rightarrow WW$ 
also very small, the amplitude being similar to that of neutrinoless double-$\bg$ decay. However, as 
$W$-pair production might also proceed via the exchange of heavy Majorana neutrinos, this opens 
another channel worth investigating; indeed the process $ee \rightarrow WW$ was considered in the 
context of a search for heavy Majorana neutrinos in ref.\,\ct{greub:1996,banerjee:2015}. These authors 
also studied possible background from jet physics in hadronic $W$- and $Z$-decays.

Moreover, as the total cross section grows as $s^2/M^4$ very high energies help. For example, 
at a CM energy of $\sqrt{s} = 8$ TeV the enhancement by the energy factor amounts to $10^{8}$,  
increasing by $10^4$ for every factor of 10 in $\sqrt{s}$. As it is 
a lepton-number violating process there is no competition from other scattering processes, provided 
the lepton number of the final state can be reconstructed. The final state observed originates from 
the direct decay of the $W$-bosons. If they both decay hadronically the lepton-number violation
is $\Del L = 2$, if one or both of them decay into leptons the observable lepton-number changes by 
$\Del L = 1$ or $\Del L = 0$, but there will be missing energy because of the neutrinos from $W$-decay. 

Which component of the mass matrix $\sfm_{Nij}$ is measured by the $W$-pair production 
depends on the charged leptons in the initial state. In this respect the process has more channels 
available than the neutrinoless double-$\bg$ decay experiments. 
\vs{3}

\nit
{\bf Acknowledgement} \\
The author is indebted to prof.\ Maarten de Jong (Nikhef/Leiden University) for stimulating discussions. 
\vs{3}

\end{document}